\documentclass[twocolumn,pre,a4paper]{revtex4}
\usepackage[dvips]{graphicx}

\usepackage{amsmath}

\newcommand{\var}[1]{\mathrm{Var}\left( #1 \right)}
\newcommand{\avg}[1]{\left< #1 \right>}
\newcommand{\fZ}{\mathcal{Z}}

\newcommand{\fU}{\mathcal{U}}
\newcommand{\fC}{\mathcal{C}}

\newcommand{\fQ}{\mathcal{Q}}

\newcommand{\amu}{\bar{\mu}}
\newcommand{\achi}{\bar{\chi}}

\newcommand{\pow}{\!\cdot\!10^}

\begin{document}

\title{The critical behaviour of the Ising model on the 4-dimensional lattice}
\date{\today}


\author{P. H. Lundow}
\email[Email: ]{phl@kth.se}
\affiliation{
  Department of Theoretical Physics, 
  KTH, 
  SE-106 91 Stockholm, 
  Sweden
}

\author{K. Markstr\"om}
\email[Email: ]{klas.markstrom@math.umu.se}
\affiliation{
  Department of Mathematics and Mathematical Statistics, 
  Ume\aa{} University, 
  SE-901 87 Ume\aa, 
  Sweden
}

\begin{abstract}
    In this paper we investigate the nature of the singularity of the
    Ising model of the 4-dimensional cubic lattice. It is rigorously
    known that the specific heat has critical exponent $\alpha=0$ but 
    a non-rigorous field-theory argument predicts an unbounded
    specific heat with a logarithmic singularity at $T_{c}$.
    
    We find that within the given accuracy the canonical ensemble data
    is consistent both with a logarithmic singularity and a bounded
    specific heat, but that the micro-canonical ensemble lends
    stronger support to a bounded specific heat. 
    
    Our conclusion is that either much larger system sizes are needed
    for Monte Carlo studies of this model in four dimensions or the
    field theory prediction of a logarithmic singularity is wrong.
\end{abstract}

\maketitle
\section{Introduction}
In dimension $D\geq 5$ it is known from \cite{aizenman:81,aizenman:82}
that the Ising model on the cubic lattice exhibits mean-field critical
exponents at the critical temperature. Even earlier it was shown \cite{sokal:79}
that the specific heat obeys the mean-field exponent $\alpha=0$ for
$D\geq 4$, and that for $D\geq 5$ the specific heat is in fact bounded 
at the critical point. For $D=4$ the rigorous results which
determine that $\alpha=0$ are not strong enough to show that the
specific heat is bounded. In fact methods from field and
renormalization theory predict that the specific heat should diverge 
as $(\ln|T-T_{c}|)^{1/3}$ but this has not been possible to prove
rigorously. There are thus, at least, two possibilities here, either
the specific heat is bounded in $D=4$ as well or it diverges
logarithmically.

Earlier studies of the critical behaviour in 4-dimensions include
\cite{BJM:02,blote:80,SV:87}, using Monte Carlo methods, and
\cite{hellmund:06}, using series expansion and extrapolation.  There
has also been some recent controversy
\cite{Cea:hep-lat0501013,st:05,BN:06} regarding the consistency of
field theoretical predictions and Monte Carlo data.

Using a standard Monte Carlo approach to detect a divergence of the
form $(\ln|T-T_{c}|)^{1/3}$ is difficult since the quantity will remain
quite small for a large range of the lattice size $L$, thereby making
it difficult to use sampled data to clearly distinguish between
different asymptotic behaviours. 

In an attempt to get around this problem we have instead studied the
microcanonical density of states of the model, following the methods
used in e.g. \cite{sampart,cubeart}. The finite-size effects of the
canonical ensemble have two components; that coming from the fact that
only a certain discrete set of energies are available in finite discrete
systems, and that coming from finite-size effects of the density of
states. The microcanonical ensemble is affected by only the latter effect.

A divergence in the specific heat means that the second derivative of
the density of states must become 0 at the critical point.  The
surprising simulation result is that this value is in fact
\emph{increasing} with the lattice size at the critical point and the
best fit to the data is that it converges to a non-zero value, thereby
also giving a bounded specific heat in the limit.

In order to make sure that this was not an artifact caused by our
simulation software we wrote two separate programs, one for the
Metropolis algorithm and one using the Wolff-cluster algorithm
\cite{wolff:89}, to sample at interleaving lattice sizes, but no
systematic differences could be seen. We also
tried to push the simulations to large lattices, reaching $L=80$. Our 
simulations give estimates for the critical exponents which agree well
with the rigorous mean-field values and a value for the critical temperature
which agrees well with earlier studies.

Hence our conclusion is that either lattice sizes larger than $L=80$
are needed to see the asymptotic behaviour of the specific heat or the
specific heat is in fact bounded at the critical point. Finding a way to
settle this issue is of prime importance since, as discussed in e.g. 
\cite{BN:06}, this would have consequences for the renormalization
techniques used to bound the Higgs mass.

\section{Notation and basic definitions}\label{sec:notation}
The lattice studied here is the cartesian graph product of
four $L$-cycles, that is, an $L \times L \times L \times L$-lattice
with periodic boundary conditions on $n=L^4$ vertices and $m=4\,L^4$ edges.  
We have collected sampled data using the sampling scheme described in 
\cite{sampart} for linear orders: $L=4, 6, 8, 10, 12, 16, 20, 24, 32, 40,
48, 56, 60, 64, 80$. For most orders we used the Metropolis
single-spin flip method with measurements of local energies after
every sweep. Since the flip-rate near the critical temperature is
about 63\% there will be no strong dependency between
measurements of local energies. For comparison we also employed the Wolff-cluster 
method for the cases $L=10, 20, 40, 60$, flipping clusters until an expected
$L^4$ spins were flipped.

The energy $E$ of a state $\sigma=(\sigma_1,\ldots,\sigma_n)$, with
$\sigma_i=\pm 1$, is defined as $E(\sigma) = \sum_{\{i,j\}}
\sigma_i\,\sigma_j$, with the sum taken over all the edges $\{i,j\}$,
and the magnetisation $M$ is defined as $M(\sigma)=\sum_i\sigma_i$
with the sum taken over all the vertices.

We have two classes of quantities. First the combinatorial
quantities from the microcanonical ensemble which depend on the energy
$U$. Especially the coupling $K$ is of interest here, defined as $K(U)
= -\partial S / \partial U$ where $S(U) = (\ln a(E))/n$ for $U=E/n$
and $a(E)$ denotes the number of states $\sigma$ at energy $E$.  How
to obtain the coupling from sampled data is described in detail
in \cite{sampart} and error estimation in \cite{recon}.

The canonical, or physical, quantities are obtained as cumulants, or
derivatives of $\ln \fZ(K,H)$ with respect to $K$ or $H$ (the
external field), where $\fZ$ is the partition function. All quantities
are measured with the external field switched off, ie $H=0$ after the
relevant derivative is taken.

At this point we introduce the notation $c_i=\avg{(X-\avg{X})^i}$ for
the $i$th central moment of a random variable $X$, where $\avg X$ is
the mean value.  The $k$th cumulant of $E$ is then the $k$th
derivative of $\ln\fZ$ with respect to $K$, where $\fZ$ is the
partition function.  Recall that the first cumulant is $\avg X$, the
second is $c_2(X)=\var{X}$, the third is $c_3(X)$ and the fourth is
$c_4(X)-3\,c^2_2(X)$. The internal energy is then $\fU(K)=\avg{E}/n$
and the specific heat is $\fC(K)=\var{E}/n$.  Note also that the
susceptibility $\chi = \var{M}/n=\avg{M^2}/n$ has no local maximum,
whereas the (spontaneous) susceptibility $\achi = \var{|M|}/n$ does.
Analogously we define the magnetisation as $\mu=\avg{M}/n$ and the
spontaneous magnetisation as $\amu=\avg{|M|}/n$.

\section{Physical quantities}\label{sec:physical}

\begin{figure}[!hbt] 
  \begin{center}
    \includegraphics[width=0.49\textwidth]{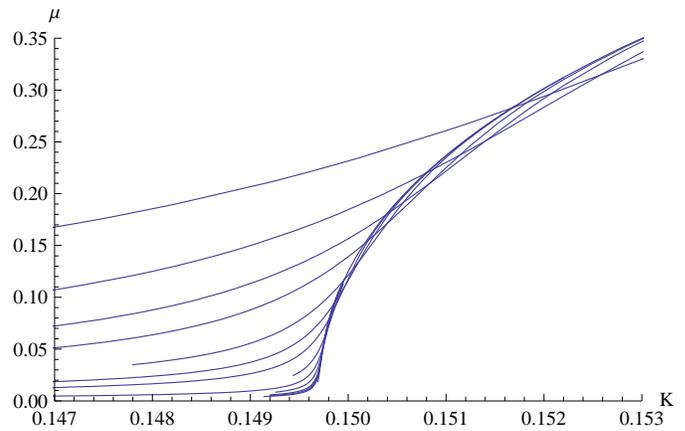} 
  \end{center} 
  \caption{(Colour online) \label{fig:M}Magnetisation $\amu(K)$ for
    lattice sizes $L=6, 8, 10, 12, 16, 20, 24, 32, 40, 48, 56, 60, 64,
    80$.}
\end{figure}

\begin{figure}[!hbt]
  \begin{center}
    \includegraphics[width=0.49\textwidth]{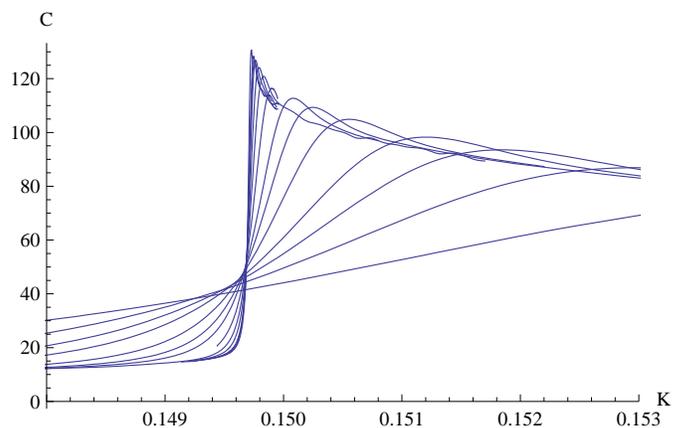}
  \end{center}
  \caption{(Colour online) \label{fig:C}Specific heat $\fC(K)$ for
    lattice sizes $L=6, 8, 10, 12, 16, 20, 24, 32, 40, 48, 56, 60, 64,
    80$.}
\end{figure}

Let us begin by showing some plots of a few physical quantities near
the critical coupling. Figure~\ref{fig:M} shows the magnetisation
$\amu(K)$. In Figure~\ref{fig:C} we show the specific heat
$\fC(K)$ for several lattice sizes.

\subsection{Critical points and exponents}\label{sec:phys-points}
First we establish a high-precision estimate of the critical coupling
$K_c$.  This is done by determining the critical points for a number
of different quantities, listed below, for each system size.  The
critical points in question are, with one exception, the locations of
various maxima or minima of eg cumulants.  To these points we fitted a
simple scaling law of the form $c_0+c_1\,L^{-\lambda}$.  By selecting
points for $L\ge L_{\min}$ for different $L_{\min}$ we can then obtain
several (for $L_{\min}=24, 32, 40$, with a few exceptions) different
estimates of the fitting parameters.  As a rule we received very good
fits deeming a higher order correction term unnecessary.  The sought
parameter is of course $c_0$.  Taking the median of these gives a
final estimate of $K_c$ for that particular quantity.  Repeating this
for all quantities, a grand total of 15, allows us to make a statistical 
analysis of them.  We have used the median as the estimate, with the first 
and third quartile as error estimates. In short, we take the median of the 
medians, very much like in~\cite{cubeart}.

\begin{figure}[!hbt]
  \begin{center}
    \includegraphics[width=0.49\textwidth]{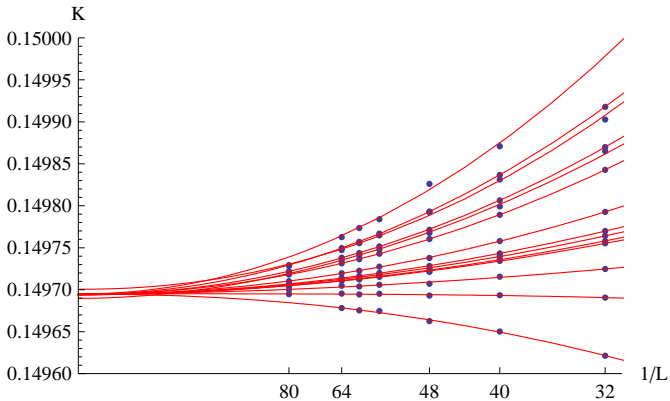}
  \end{center}
  \caption{(Colour online) \label{fig:Kstar}The critical points vs
    $1/L$ with fitted curves.}
\end{figure}

The points scale very nicely with the linear order using only the simple 
expression above, see Figure~\ref{fig:Kstar}. The
resulting estimate is $K_c = 0.1496947 \pm 5\pow{-7}$. This falls
inside the by now rather old estimate $K_c=0.14965 \pm 5\pow{-5}$
found in~\cite{gaunt:79} and agrees with the estimate from
\cite{BJM:02}.

The critical points in question are the locations of the following;
the maximum of the specific heat $\fC$ and susceptibility $\achi$,
maximum and minimum of the cumulants $c_3(E)/n$, $c_3(|M|)/n$,
$(c_4(E)-3\,c_2(E))/n$ and $(c_4(|M|)-3\,c^2_2(|M|))/n$, maximum of
$\partial\amu/\partial K$, $\partial\ln\amu/\partial K$,
$\partial\ln\chi/\partial K$ and $\partial\fQ/\partial K$, where
$\fQ$ is the Binder cumulant $1-\avg{M^4}/3\,\avg{M^2}^2$ and finally
the crossing point between $\fQ_L$ and $\fQ_{L/2}$. See eg
\cite{ferrenberg:91} for a discussion of the last four quantities.

The expression above also provides us with estimates of the exponent
$\nu$.  The location of a critical point $K^*_L$ should deviate from
$K_c$ as roughly $K^*_L-K_c \sim L^{-1/\nu}$, again
see~\cite{ferrenberg:91}.  Repeating the median-of-the-medians
approach gives $\lambda = 1/\nu = 2.00 \pm 0.03$ where the bounds are
again based on the 1st and 3rd quartile, thus rendering us $\nu = 0.50
\pm 0.01$.  The Josephson inequality tells us that $\alpha \ge
2-D\nu$, and hence our midpoint estimate gives $\alpha \ge 0 \pm 0.04$
for $D=4$, since $\alpha=0$ \cite{sokal:79} our data is in good
agreement with the rigorous results. Similarly an estimate of
$\beta=0.50$ is found, and the mean-field value is $\beta=\frac{1}{2}$.

Having established an estimate of $K_c$ we can now estimate the
internal energy $\fU_L(K_c)$ and again fit the scaling formula above
to these data for $L_{\min}=24,32,40$.  The different $c_0$, and thus
the asymptotic values of $\fU_c$, end up inside the interval
$0.77053\pm 4\pow{-5}$.
  
\subsection{Critical values}\label{sec:phys-values}
Our aim is now to try to distinguish between the two possible
scenarios, either we have a logarithmic singularity or the specific
heat is bounded at $T_{c}$. We attempt to do this by making
least-squares fits to the data for two different forms of the fitting
function.

\begin{figure}[!hbt]
  \begin{center}
    \includegraphics[width=0.49\textwidth]{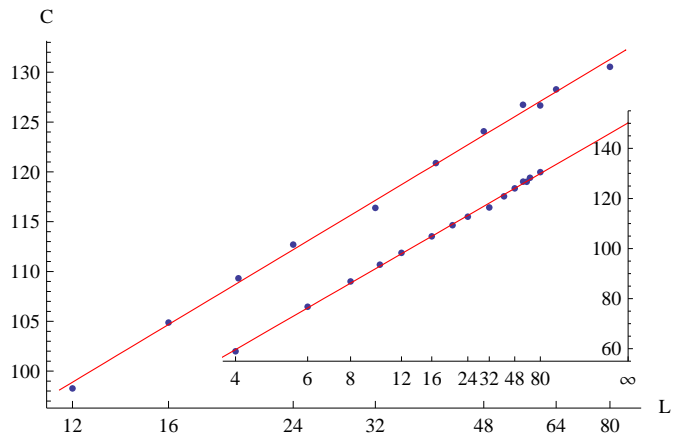}
  \end{center}
  \caption{(Colour online) \label{fig:Cmax}$\fC_{\max}$ vs
    $\sqrt[3]{\ln L}$ and $L^{-0.496}$ (inset), together with the
    fitted curves.}
\end{figure}

According to scaling theory, see~\cite{ptcp}, the maximum specific
heat $\fC_{\max}$ is proportional to $\sqrt[3]{\ln L}$.   For $L\ge 12$
this seems plausible given our data. In
Figure~\ref{fig:Cmax} we show $\fC_{\max}$ versus $\sqrt[3]{\ln L}$
together with a fitted straight line, $y=115\,x-56.7$, and indeed they
line up rather convincingly. The reader should
note that $\fC_{\max}$ grows very slowly indeed. 

For the bounded scenario we try a fit where $\fC_{\max}$ is
proportional to a power of $L$. A least-squares fit of both constant
and exponent gives $150.49+180.5L^{-0.496}$. We show this in the inset of 
Figure~\ref{fig:Cmax}. The fact that the exponent is negative 
would of course mean that the specific heat  is finite in the limit.

For both models there is some variation in the coefficients and the
exponent if one makes the fit to different subsets of the data points,
but no drastic changes.  An attempt with evaluating the specific heat and the
susceptibility at the asymptotic $K_c$ for each linear size instead 
gave a very similar behaviour to that of their maximum value.

To the eye both fitting functions work reasonably well and we simply find that 
the canonical ensemble data can not strongly distinguish the two
scenarios.

\section{Combinatorial quantities}\label{sec:combinatorial}
With regards to the microcanonical ensemble the two scenarios will be that
either $K'(U)$ goes to 0 at $U_{c}$ or it converges to a finite
positive value

\begin{figure}[!hbt]
  \begin{center}
    \includegraphics[width=0.49\textwidth]{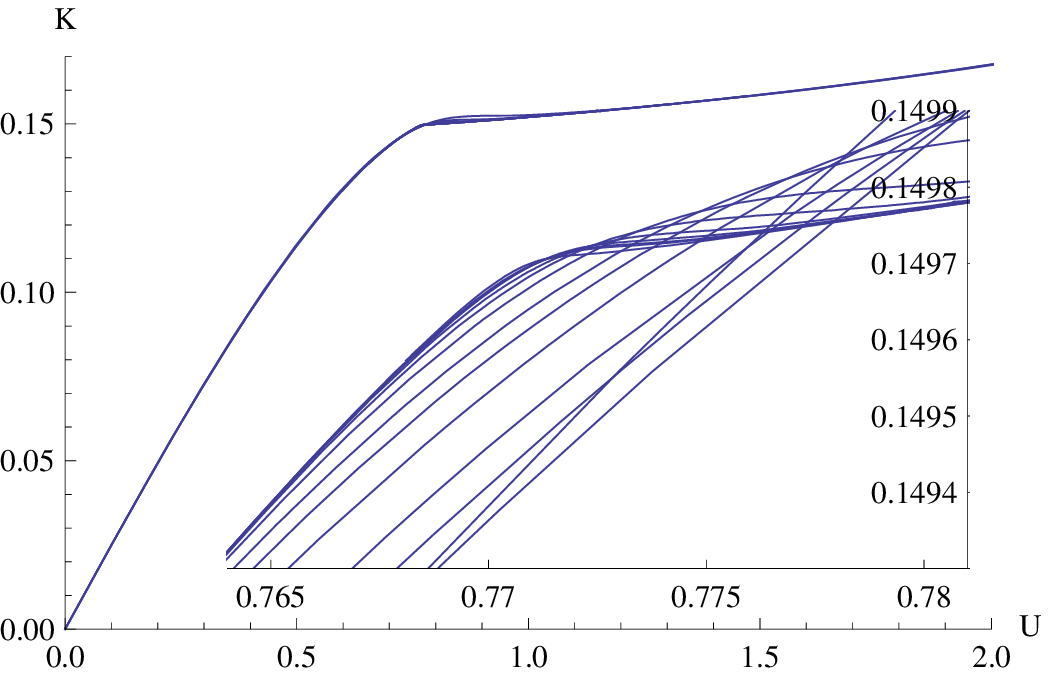}
  \end{center}
  \caption{(Colour online) \label{fig:K}Coupling $K(U)$ for $L\ge 6$.}
\end{figure}

\begin{figure}[!hbt]
  \begin{center}
    \includegraphics[width=0.49\textwidth]{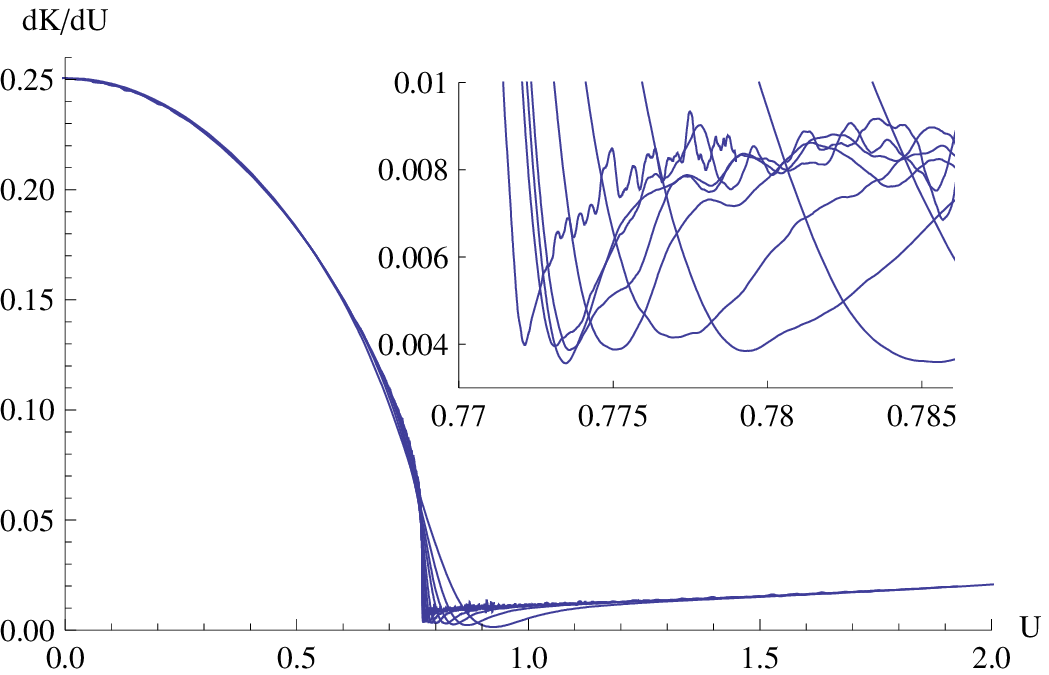}
  \end{center}
  \caption{(Colour online) \label{fig:Kd}Coupling $K'(U)$ for $L\ge 6$.}
\end{figure}

Figure~\ref{fig:K} shows the microcanonical quantity $K(U)$ and in
Figure~\ref{fig:Kd} its derivative is shown, both together with
zoomed-in versions near the critical energy $U_c$.  Most of the
sampling was done for energies close to the critical one for the given 
value of $L$ so the curves become noisier further away from $U_{c}$.

The minima do not at all seem to approach zero as they do for
$d=2$~\cite{exactart} and $d=3$~\cite{cubeart}. In fact the behaviour
here is qualitatively different in that the values are actually
increasing rather than decreasing.

\begin{figure}[!hbt]
  \begin{center}
    \includegraphics[width=0.49\textwidth]{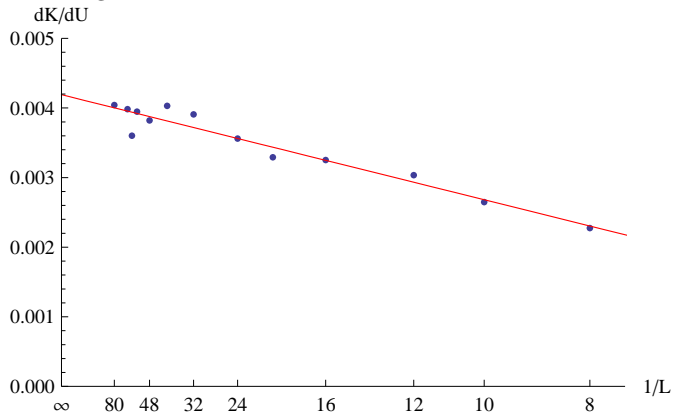}
  \end{center}
  \caption{(Colour online) \label{fig:Kdmin}The minimum of $K'(U)$ vs
    $1/L$, together with the fitted curve.}
\end{figure}

It is  known, see e.g.  \cite{sampart}, that the
specific heat corresponds to $1/K'(U)$.  Thus $\lim_{U\to U_c} K'(U) =
0$ if and only if $\lim_{K\to K_c} \fC(K) = \infty$.
Figure~\ref{fig:Kdmin} shows the minima versus $1/L$ together with a
fitted line $y=0.00419-0.0151\,x$, suggesting that the minimum
approaches a maximum $0.00419$.  

The optimal exponent of $1/L$, naturally, depends to some extent on
which data points are used.  Using a least-squares fit to different
subsets of the data for $L\ge 6$ gives exponents between (roughly)
$0.9$ and $1.5$.  More specifically, if we check all subsets of the data
with $L\ge 6$ on between 10 and 12 points a median exponent of $1.25$
is received and for $c_0$ the median value was $0.00406$ with first
and third quartiles $0.00402$ and $0.00413$ respectively. The extremal
values for $c_{0}$ are $0.0038$ and $0.0044$.  If we
instead use all the data points for $L\ge 8$ we obtain the exponent
$\lambda=1.147$ and $c_0=0.00411$.

\section{Conclusions}
We have studied the two proposed scenarios for the critical behaviour
of the specific heat of the 4-dimensional Ising model.  This has been
done in both the canonical and the microcanonical ensembles.  We have
found that for the given lattice sizes the canonical ensemble can not
conclusively distinguish between the two scenarios, and in an attempt
to circumvent this we have instead turned to the microcanonical ensemble.

There are two reasons for why the microcanonical ensemble could give
clearer results in this situation, the first predicted and the second 
unexpected.

First, the canonical ensemble is
expected to have larger finite size effects than the microcanonical
ensemble. To see this we may consider an idealised example where, for a
finite system,  $S(U)$ at each energy $U$ is identical to the limit as
$n\rightarrow \infty$. Here the density of states has no finite size
effects at all, apart from only being defined at certain discrete set of
values of $U$. However because of the discrete energies there will
still be finite-size effects in the corresponding canonical ensemble.

Secondly, a divergent specific heat means that $K'(U)$ goes to 0 at
$U_{c}$, and as we have found the minimum value of $K'(U)$ is actually
increasing rather than decreasing. This gives us a qualitative
signal, rather than a weak quantitative one, that the specific heat
actually converges to a finite value. 
 
Our conclusion is that either much larger systems are needed to see
the asymptotic behaviour of this model, and this possibility can only be 
ruled out by a rigorous convergence result, or the specific heat is in
fact bounded at $U_{c}$, thus contradicting the renormalization
theory prediction.

\section*{Acknowledgements}
This research was conducted using the resources of High Performance
Computing Center North (HPC2N) and the Center for Parallel Computers (PDC).
Thanks are also due to the referees for their constructive criticisms.


\end{document}